\documentclass[pss]{wiley2sp} 
\usepackage{amsmath}

\begin{document}

\title{Scanning Tunneling Microscopy and Spectroscopy of Graphene on Insulating Substrates}

\titlerunning{STM Graphene}

\author{%
  Markus Morgenstern\textsuperscript{\Ast,\textsf{\bfseries 1}},
  }

\authorrunning{M. Morgenstern}

\mail{e-mail
  \textsf{mmorgens@physik.rwth-aachen.de}, Phone:
  +49-241-8027075, Fax: +49-241-8022306}

\institute{%
  \textsuperscript{1}\,II. Institute of Physics and JARA-FIT, RWTH Aachen, 52074 Aachen, Germany}

\received{XXXX, revised XXXX, accepted XXXX} 
\published{XXXX} 

\keywords{Scanning tunneling microscopy, graphene, nanomechanics, Landau levels}

\abstract{%
%
%
%
\abstcol{%
  Graphene is a truly two-dimensional material with exceptional electronic, mechanical, and optical properties. As such, it consists of surface only and can be probed by the well developed surface-science techniques as, e.g., scanning tunneling microscopy.
  This method bridges the gap between the surface science community and the electronic device community and might lead to novel combined
  approaches.
  }{%
  Here, I review some of the scanning tunneling microscopy (STM) and spectroscopy (STS) experiments on monolayer graphene samples. I will concentrate on graphene samples deposited on insulating substrates, since these are related to graphene device concepts.  In particular, I will discuss the morphology of graphene on SiO$_2$ and other emerging substrates, some nanomechanical manipulation experiments using STM, and spectroscopic results. The latter can map the disorder potentials as well as the interaction of the electrons with the disorder which is most pronounced in the quantum Hall regime.}}

%
%
\titlefigure[height=6.1cm]{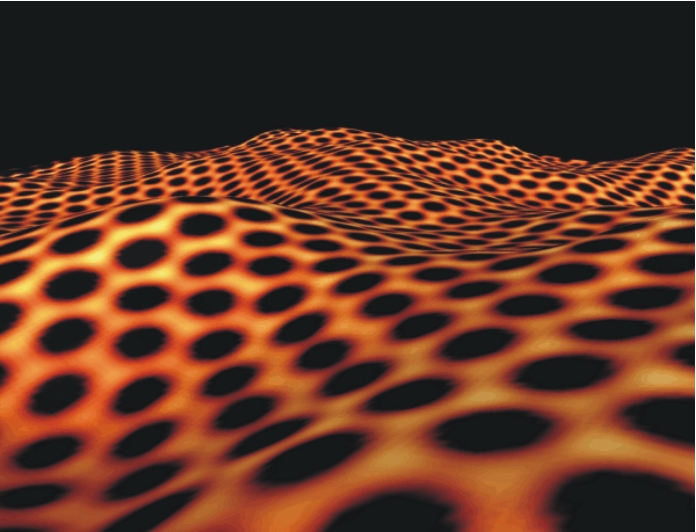}
\titlefigurecaption{%
  3D-representation of a scanning tunneling microscopy image of graphene; the atomic resolution is displayed
  as color code, while the rippling is shown in 3D. (courtesy of M. Pratzer, RWTH Aachen).}

\maketitle   

\section{Introduction}
Since its discovery in 2004 \cite{Geim}, Graphene has become one of the most investigated materials worldwide with more than 3500 publications per year. This is partly due to the exceptional band structure of graphene which might mimic Dirac physics in table-top experiments \cite{Geim2,Castro}, but increasingly due to the applicational prospects with respect to, e.g., transparent electrodes \cite{Samsung}, high frequency transistors \cite{Avouris}, optoelectronic devices \cite{Ferrari} and sequencers for individual DNA strands \cite{Kouwenhoven}. Further applications in spintronics \cite{vanWees,Beschoten} or quantum computation \cite{Ensslin} are envisioned.\\
Particular advantages of graphene are the abundance of its constituent atoms, its inertness, its very high
breaking strength, its high room-temperture mobility, and its simple preparation. Moreover, as a 2D material it consists of surface only and can be characterized in depth by the well developed surface science techniques such as angular resolved
photoelectron spectroscopy (ARPES) \cite{Seyller} or scanning tunneling microscopy (STM). The latter technique reveals precise information on morphology and electronic structure. In short, graphene offers the unique possibility to combine precise knowledge of a device down to the atomic scale with functionality. This feature article gives a short summary of the major achievements obtained with
STM so far. It concentrates on graphene on insulators, since metal substrates are limited with respect to electronic applications due to the shortcircuit created by the metal. As usual, only a selection of results can be presented.\\

\section{Scanning tunneling microscopy}
In scanning tunneling microscopy, a sharp metallic tip, mostly ending in a single atom, is positioned $3-8$ ${\rm \AA}$ above a
conducting surface. The surface is usually prepared in ultra high vacuum (UHV) in order to be atomically smooth. The high inertness of graphene allows preparation outside UHV. Subsequent annealing to, e.g., 400$^\circ$ C, is sufficient to get rid of all unwanted adsorbates leaving an atomically smooth graphene surface as presented, e.g., in the introductory figure. A voltage $V$ is applied between the tip and the conducting surface and the resulting tunneling current $I$ is measured. $I$ depends exponentially on the distance between surface and tip $\Delta z$ according to $I(z) \propto e^{-\alpha \Delta z}$. A good approximation for $\alpha$ is
$\alpha = \sqrt{(4m \cdot (\Phi_s+\Phi_t -e|V|)}/\hbar)$ with Planck's constant $\hbar$, electron mass $m$, electron charge $e$ and  work functions of tip and sample $\Phi_t$ and $\Phi_s$, respectively. A good estimate is $\alpha \simeq 2.1$/{{\rm \AA}}.\\
The tip is positioned with respect to the sample using piezoelectric elements. All three directions $x$, $y$, and $z$ can be changed with sub-pm precision \cite{Mashoff}. For STM, the tip is scanned in $x$ and $y$ direction and the tunneling current is kept constant by a feedback mechanism adjusting $z$. The resulting $z(x,y)$ is plotted and called constant-current image. It represents,
to first order, a contour of constant integrated local density of states  of the substrate $LDOS(x,y,z,E)$, where the integration has to be taken between the Fermi levels of sample and tip to be adjusted by $V$ \cite{Morgenstern3}. The central position of the very last atom of the tip is given by $(x,y,z)$ and $E$ is the energy. Such images are often called topography of the sample indicating that corrugations of the atomic positions dominate the contour.\\
Differentiating $I$ with respect to $V$ (at low $V$ with respect to $\Phi_{s(t)}$) gives direct access to the $LDOS$ according to:
\begin{eqnarray}
dI/dV(x,y,z,V) \propto LDOS(x,y,z,E) =\\ \nonumber
\sum |\Psi_{\tilde{E}}(x,y,z)|^2\cdot \delta(E-\tilde{E}).
\end{eqnarray}
Thereby, $\Psi_{\tilde{E}}$ are the single-particle wave functions of the substrate at energy $\tilde{E}$ and $E=eV$.
Of course, this requires that the system is adequately described by independent single-particle wave functions. Moreover, an s-type
symmetry of the orbitals of the last atom is necessary, in principle, but Chen has shown that the model remains largely correct even
if higher orbital momenta are contributing \cite{Chen}.
In real experiments, the $\delta$-function has to be replaced by an energy resolution function with approximate full width of half-maximum of $\delta E\approx\sqrt{(3.3\cdot k_BT)^2+(1.8\cdot eV_{\rm mod})^2}$. $T$ is the temperature ($k$: Boltzmann's constant) and
$V_{\rm mod}$ is a modulation amplitude used to detect $dI/dV$ by lock-in technique. The resulting $dI/dV(x,y)$ recorded at constant $V$ and $z(x,y)$-values resulting from constant $I$ is called $LDOS$-image.\\
Thus, STM can measure atomic structure with sub-pm resolution and electronic structure ($LDOS$) with sub-meV resolution \cite{Mashoff}. The energy resolution makes STM complementary to the transmission electron microscope (TEM) which reveals atomically well defined structural information partly with chemical specifity \cite{Meyer}, but not the $LDOS$ down to the meV scale.
\subsection{Graphene}
Graphene is a two-dimensional (2D) honeycomb lattice of carbon atoms, which are bound in sp$_2$+p$_z$ configuration leading to
$\sigma$-bonds and $\pi$-bonds, respectively \cite{Wallace}. As a 2D material, it should be unstable at $T>0$ K \cite{Peierls}, but it has been found theoretically that a coupling of out-of-plane and in-plane phonons could stabilize graphene without a support leading to a rippled morphology \cite{Katsnelson}. Such rippling with wave length of about 10 nm and amplitude of about 1 nm has indeed been found experimentally, if the graphene is only partly supported at the edges, e.g. around a hole of a TEM grid, and freely suspended in between \cite{Meyer2}. It is still debated, if the experimentally found rippling is fundamental or if it is caused by preparation history \cite{Locatelli}.\\
\begin{figure*}[thb]
\centering{\includegraphics*[width=0.9\textwidth]{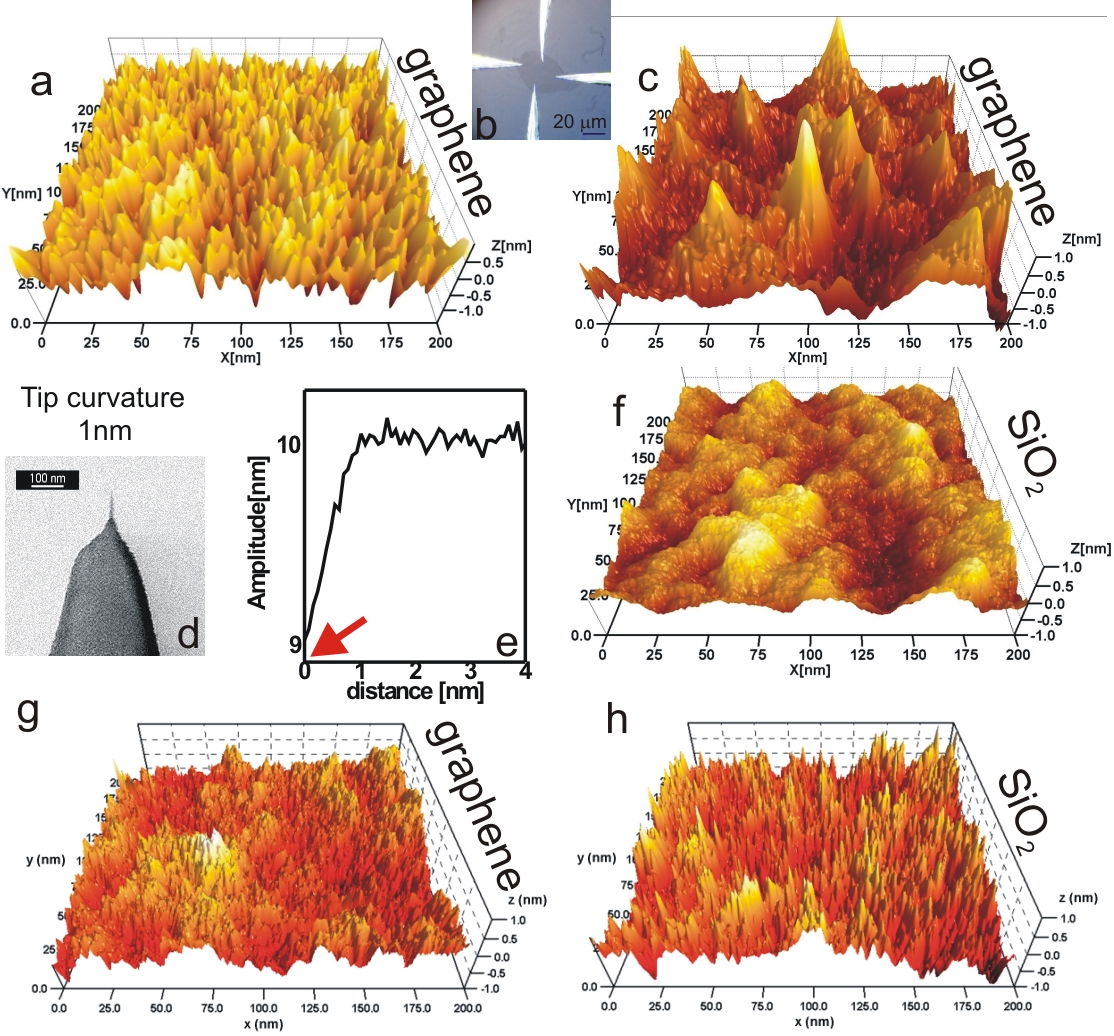}
\caption{(a) STM image of graphene deposited on SiO$_2$ by the scotch tape method, $V= 1$ V, $I=0.2$ nA, $T=300$ K; (b) optical microscope image of a monolayer graphene sample contacted by In microsoldering \cite{Girit,Geringer2}; (c) same as (a), but on another sample, $V=0.4$ V, $I=0.2$ nA, $T=300$ K;
 (d) scanning electron microscopy image of the W tip used for the AFM measurements shown in (f), nominal curvature radius at the apex is indicated; (e) oscillation amplitude of the cantilever as a function of tip-sample-distance, the tip-sample distance used for recording (f) is marked by an arrow; (f) Tapping-mode AFM-image of the SiO$_2$ substrate used for the preparation of (a) and (b), $\Delta f=-100$ Hz, force constant: 47 N/m, oscillation amplitude: $A=18$ nm \protect{\cite{Geringer}}; (g) same as (a) and (c), but measured by another group, $V=-0.3$ V, $I=0.04$ nA, $T=300$ K; (h) non-contact AFM image of the SiO$_2$ used for the preparation of (f), $\Delta f=-20$ Hz, $A=5$ nm
\protect{\cite{Cullen}}; all images execept (f) are recorded in ultrahigh vacuum. (courtesy of W. G. Cullen, University of Maryland for (g),(h))}
\label{Fig1}}
\end{figure*}
The electronic structure of graphene at $E_F$ is governed by the bonding and antibonding parts of the $\pi$-band. These bands touch at the corners of the hexagonal Brillouin zone (BZ) due to the fact that the unit cell exhibits two atoms, which are attributed to two different sublattices. The energy dispersion $E(|\underline{k}|)$ is, to very good approximation, linear around $E_F$ leading to conical $E(\underline{k})$ planes: $E(\underline{k})= \hbar v_D |\underline{k}|$ ($\underline{k}$: wave vector, $v_D\simeq 10^6$ m/s: Dirac velocity). In first order, the bands are also electron-hole symmetric, i.e. a cone is opening upwards into the unoccupied states and downwards into the occupied states from the six points at the corners of the BZ. The energy at the crossing points is called Dirac
energy $E_D$ since it can be moved away from $E_F$ by charging the graphene layer, either using dopants or a gate voltage.\\
In addition, the electronic wave functions show a sublattice texture, which changes continuously, if one moves around the cone \cite{Castro}. Since the sublattice textures at opposite sides of the cone are orthogonal, backscattering of electrons is suppressed
\cite{Ando}. This is regarded as a major reason for the high electron mobility of up to $\mu \simeq 10$ m$^2$/Vs found close to room temperature \cite{Bolotin} (for comparison: mobility of Si at 300 K: $\mu \simeq 0.1$ m$^2$/Vs). The linear band structure and the sublattice texture renders the electrons in graphene equivalent to chiral massless particles at the velocity of light albeit with the reduced velocity $v_D$. Thus, graphene raises hope that relativistic effects as Zitterbewegung, Klein tunneling or supercritical charge
might be observed in table-top experiments \cite{Castro,Pereira}.\\
In magnetic $\underline{B}$-field, the linear dispersion of graphene is divided into Landau and spin levels. The energy of the Landau levels is given by $E_N = \pm v_{\rm D} \cdot \sqrt{2e\hbar\cdot |N|B}$ ($N\in Z$). The square-root dependence on $|\underline{B}|$ leads to a relatively large separation of the central Landau level $E_0$, which, protected by symmetry, is located at $E_D$, and the two surrounding Landau levels. An energy distance of $\Delta E =0.1$ eV can be reached already at $B=5$ T, which is partly responsible for the fact that quantum Hall plateaus can be observed at 300 K \cite{Zeitler}. Consequently, it is attractive to study quantum Hall physics, which is driven by interaction of electrons with disorder and, thus, by inhomogeneous electron distribution \cite{Prange}, using a local scale method like STM on graphene.\\
Also the more intricate fractional quantum Hall effect \cite{Prange} is significantly more stable in graphene than in the conventionally used III-V semiconductors \cite{Andrei,Deshpande3}.
This can be rationalized, since the energy gaps, which cause the additional plateaus in the Hall curve at particular ratios between electron density $n$ and density of magnetic flux units $n_{\rm LL}$, are driven by electron-electron interactions \cite{Prange}. These interactions are much more intense within a single atomic layer surrounded by a dielectric material with $\epsilon =1$ than in a 2D electron system in III-V semiconductors with thickness 10 nm and surrounded by $\epsilon=13$. The energy gaps $E_{\rm gap}$ at $\nu:=n/n_{\rm LL}=1/3$ of freely suspended graphene were indeed found to be $E_{\rm gap}=4$ meV at $B=20$ T \cite{Andrei} and even after depositing the graphene on BN(0001), they were $E_{\rm gap}=2$ meV at $B=12$ T \cite{Deshpande3}. Thus, studying fractional quantum Hall physics on the nm scale might be within reach on graphene.\\
Notice that each Landau level of graphene is fourfold degenerate due to usual spin degeneracy and due to the two non-equivalent corners of the Brillouin zone K and K', which leads to so-called valley degeneracy.
\section{Graphene Morphology}
The first graphene samples of one monolayer height have been prepared by the scotch-tape method on SiO$_2$ \cite{Geim}, but shortly after researchers also managed to show single-layer behavior for the sublimation method applied to SiC(0001) \cite{Berger}. Since the later samples cover a complete area, they are easier to find by STM and, thus, have been imaged rather early in ultra high vacuum \cite{Mallet}. It has been found that single layer graphene exhibits a $\sqrt {3} \times \sqrt{3}$ reconstruction \cite{Mallet} being located on a strongly bound carbon layer in sp$_3$ configuration \cite{Mallet3}.\\
But also graphene flakes deposited on SiO$_2$, which exhibit higher mobility and mostly better
defined quantum properties \cite{Geim2,Tikhonenko,Weber} have been imaged by STM. They exhibit a rippling on the 10 nm length scale with amplitudes of about 1 nm \cite{Stolyarova,Ishigami}. Figure \ref{Fig1}(a), (c), and (g) show three STM images of typical graphene flakes prepared by the scotch tape method in comparison with atomic force microscopy (AFM) measurements of the respective SiO$_2$ substrates as shown in Fig. \ref{Fig1}(f) and (h) \cite{Geringer,Cullen}.  Partly, the length scales between substrate and graphene corrugation coincide (Fig. \ref{Fig1}(c)/(f) and (g)/(h)), but partly the length scale of the graphene rippling is a factor of 3-4 smaller than the length scale of corrugation on SiO$_2$ (Fig. \ref{Fig1}(a)/(f)). Since the latter might be an artifact of reduced lateral resolution of the AFM measurement, the authors used a particularly sharp tip as shown in Fig. \ref{Fig1}(d) and controlled the contact depth by oscillation amplitude vs. distance curves as shown in Fig. \ref{Fig1}(e). This provides a lateral resolution of 1 nm much shorter than the observed length scale of rippling. A detailed analysis using Fourier transformation and autocorrelation functions of the real space images revealed that the latter type of samples, which are found in 80 \% of the preparations reported in \cite{Geringer}, are partly freely suspended above the substrate and exhibit a slightly preferential wave length of 15 nm, which surprisingly is very similar to the rippling found on freely suspended graphene flakes by TEM \cite{Meyer2}. However, the image within the abstract, which is taken on the same sample, shows that rippling on all length scales down to about 2 nm is present.\cite{Geringer}. \\
The other samples (Fig. \ref{Fig1}(a)/(f) and (g)/(h)) show a nearly perfect conformation of the graphene flake to the substrate . This is to be expected due to the attractive van-der-Waals forces between the substrate and the graphene outperforming
the elastic restoring forces of the graphene \cite{Cullen}. The latter are fully characterized by the elastic modulus measured to be $E=340$ N/m  \cite{EModul}.\\
Obviously, the details of the morphology of graphene on SiO$_2$ depend on the uncontrolled details of the preparation procedure, which are also apparent, e.g., by mobility values varying by a factor of 100 \cite{Geim2,Geim3}. So far, it is unclear, in how far the rippling is responsible for the disorder potential and the mobility of graphene \cite{Polini}. However, depositing graphene on an intermediate BN layer decreases the rippling down to a value of
about 0.01 nm \cite{Deshpande2} and, at the same time, reduces the disorder potential by more than a factor of 10 leading to strongly improved mobility \cite{Andrei}.\\
\section{Nanodrums}
\begin{figure*}[thb]
\includegraphics*[width=\textwidth]{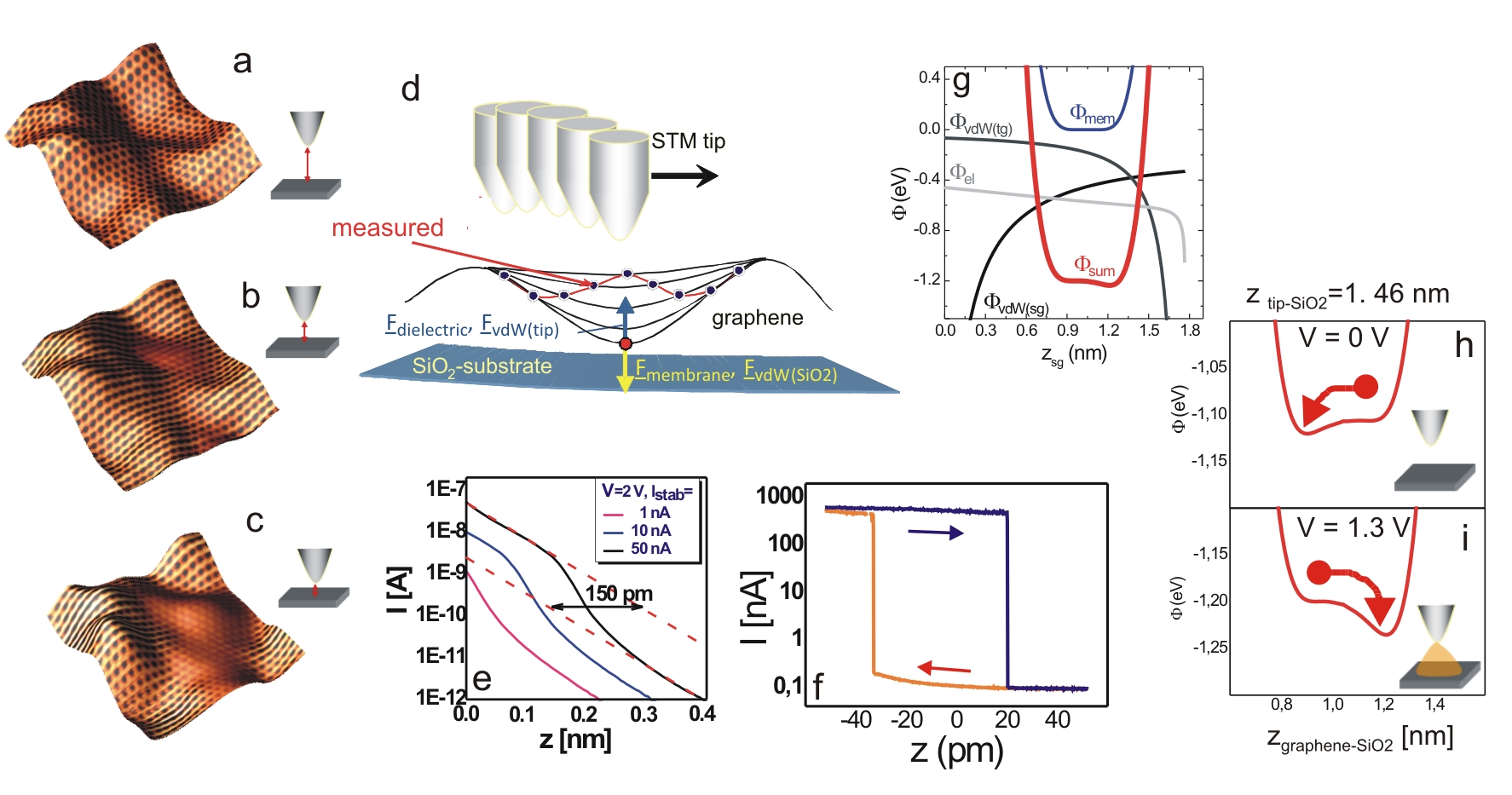}
\caption{(a) STM image of graphene deposited on SiO$_2$, $V=1$ V, $I=0.1$ nA, $T=5$ K, atomic resolution is shown as color code, while long-range rippling is displayed in 3D; (b) same as (a), but at $V=0.7$ V, $I=1$ nA; (c) same as (a), but at $V=0.4$ V, $I=2$ nA; small images indicate decreasing tip-surface distance; (d) sketch of the dynamic movement of the valley during scanning of the tip (black lines) and the resulting, measured line (red); the directions of the four forces acting on the membrane are color-coded; (e) $I(z)$ curves measured on a valley, which moves continuously towards and away from the tip; voltage $V$ and stabilization currents $I_{\rm stab}$ are indicated; the red dashed lines mark the
decay $\ln{(I/I_0)}=2\sqrt{2m\phi/\hbar}$ ($m$: electron mass, $\phi=4.85$ eV: work function averaged between tip and surface, $\hbar$: Planck's constant, $I_0$: current at saturation) expected from pure tunneling;
(f) $I(z)$ curve measured on a valley which moves hysteretically; $V=1$ V; directions of tip movement are indicated by arrows; (g) diagram showing the different potentials acting on the membrane as well as the sum potential $\Phi_{\rm sum}$; $\Phi_{\rm mem}$: elastic restoring potential of the graphene valley, $\Phi_{\rm el}$: potential induced by the image charge of the biased tip, $\Phi_{\rm vdW(tg)}$: van-der-Waals potential between tip and graphene, $\Phi_{\rm vdW(sg)}$: van-der-Waals potential between SiO$_2$ substrate and graphene; $z_{\rm sg}$: distance between graphene and substrate; (h), (i) sum potentials at the tip voltages indicated and at a tip-substrate distance of 1.46 nm; the arrows mark the movement of the membrane induced by switching the voltage between these two values.    \protect{\cite{Mashoff}}}
\label{Fig2}
\end{figure*}
The partly freely suspended graphene samples on SiO$_2$ can be used for nanomechanical manipulation of the one atomic layer thick
membrane. Therefore, the tip of the STM is used \cite{Mashoff}, whose forces on a local graphene area can overcome the van-der-Waals forces of the substrate. Figure \ref{Fig2}(a)-(c) show that valleys of the rippled graphene can indeed be lifted by the tip forces. The central valley is transformed into a hill by decreasing the distance between tip and substrate and, thereby, increasing the respective van-der-Waals force. Figure \ref{Fig2}(d) depicts that the corresponding images are dynamic, since the force between valley and tip changes during the imaging procedure. However, the I(z) curves taken within the
center of a valley can be used to classify the valleys. 10 \% of the valleys exhibit a tip-sample distance region of about 0.15 nm, which exhibits a stronger exponential increase of tunneling current than expected from the work function (Fig. \ref{Fig2}(e)). This indicates continuous lifting of the graphene valley. 50 \% of the valleys do not move at all probably because they touch the substrate. The remaining 40 \% exhibit a hysteretic $I(z)$ curve with jumps in tunneling current by about three orders of magnitude as shown in Fig. \ref{Fig2}(f). This indicates bistable behaviour of the membrane.
Indeed, adding up the elastic restoring potential of graphene and the two van-der-Waals potentials of tip and sample reveals a double well potential \cite{Mashoff} as shown in Fig. \ref{Fig2}(g). This double well potential can be additionally tilted by the dielectric force (image charge) of the biased tip. Thus, a flipping of the membrane between the two potential valleys is possible either by  applying a tip voltage as shown in Fig. \ref{Fig2}(h) and (i), or by changing the tip-sample distance and, thereby, changing the van-der-Waals force of the tip.\\
\begin{figure}[htb]
\includegraphics[width=\linewidth]{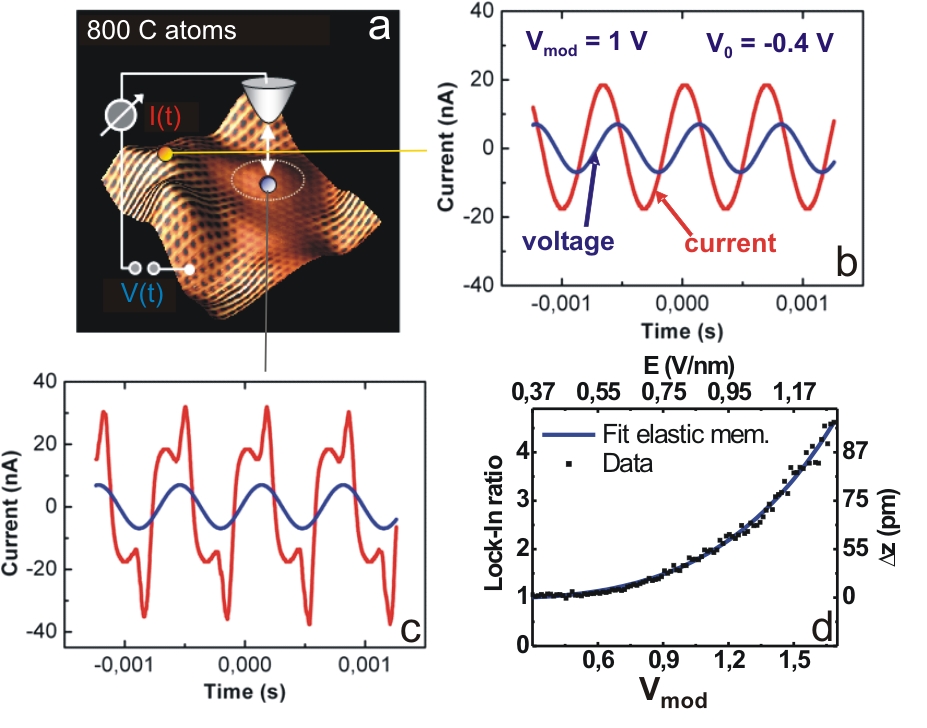}
\caption{(a) STM image of graphene deposited on SiO$_2$, $V=-0.4$ V, $I=2$ nA, $T=5$ K; tip with applied ac voltage $V(t)$ and resulting current $I(t)$ is sketched; yellow and grey dots mark measurement positions for (b) and (c), respectively; (b) applied ac voltage $V(t)=V_{\rm mod} \cdot sin{(\omega\cdot t)}$ (blue) and resulting current response $I(t)$ (red); measured at the position marked by a yellow dot in (a), tip was stabilized at $V_0=-0.4$ V, $I=2$ nA; (c) same as (b), but at the position marked by a grey dot in (a); (d) In-phase $I(t)$ amplitude at applied frequency $\omega=2\cdot \pi\cdot 1000$ rad/s measured at the grey dot in (a) as a function of voltage amplitude $V_{\rm mod}$; top and left scale show electric field amplitude $E$ and resulting oscillation amplitude of the membrane $\Delta z$ \protect{\cite{Mashoff}}.}
\label{Fig3}
\end{figure}
An interesting aspect of the movable nanomembranes is the presumably extremely large resonance frequency, which can be estimated to be 0.5 Thz \cite{Mashoff,05Thz}. It is given by the small number of contributing atoms (800), its light mass (12 u) and
the large elastic strength of graphene with an elastic modulus of $E_{\rm 2D}$=340 N/m. The corresponding vibrational energy of a movable valley is 2 meV, leading to 99.9 \%
ground state occupancy already at $T=4$ K and a complete suppression of excitations at $T=0.3$ K,  a temperature which is meanwhile achieved routinely in many STM laboratories \cite{Wiebe}.\\
In order to estimate the required electric field amplitude for exciting at resonance, we measured the oscillating mechanical response of the membrane to an oscillating voltage applied to the tip. Figure \ref{Fig3}(b) and (c) show the applied voltage and the resulting current response on a non-moving area in (b) and on a continuously moving valley in (c). The former exhibits only a 90$^\circ$ phase-shifted
sinusoidal current given by the capacitive cross-talk between tip and graphene sample. However, the latter clearly shows an additional, non-sinusoidal in-phase signal indicating the movement of the sample. The amplitude of this signal can be recalculated to a height amplitude of the oscillating membrane by using measured $I(z)$ curves on non-moving areas \cite{Mashoff}. The result is shown for different excitation amplitudes in Fig. \ref{Fig3}(d) revealing that an amplitude of 1 V/nm is required for an oscillation amplitude of 0.1 nm at the excitation frequency of 1000 Hz being far below resonance.\\
Taking the quality factor of 10$^5$ recently obtained on larger graphene flakes at much larger amplitude \cite{Bachthold}, the required voltage amplitude would be reduced to 10$^4$V/m, which should be taken as an
orientational value for future experiments heading for a resonant coupling to these nanomembranes.

\section{Scanning Tunneling Spectroscopy}
\begin{figure}[htb]
\includegraphics[width=\linewidth]{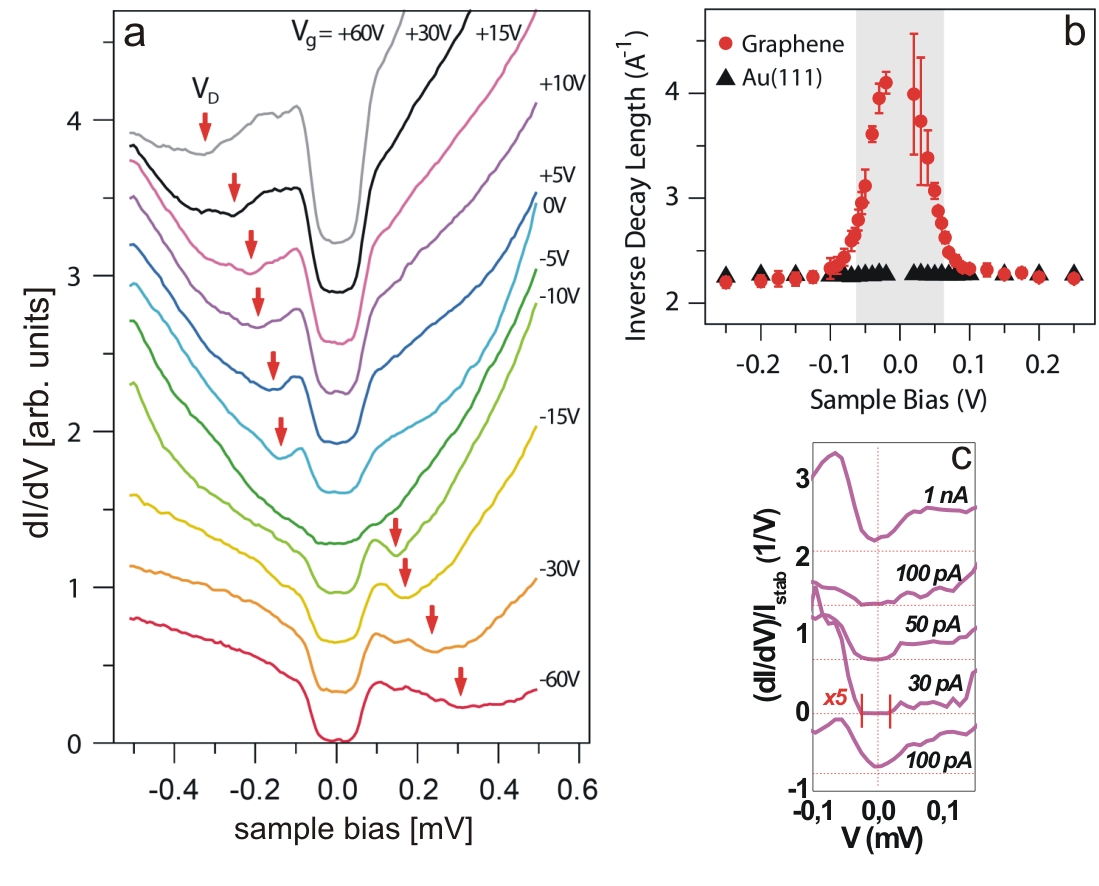}
\caption{(a) $dI/dV$ curves recorded on the same position of graphene deposited on SiO$_2$ at different back gate voltage $V_{\rm gate}$,
$V_{\rm gate}$ is given on the right of the curves, the arrow marks the Dirac point, $V_{\rm stab}=0.5$ V, $I_{\rm stab}=0.1$ nA, $T=4.8$ K; (b) inverse decay length $1/\lambda$ deduced from fitting $I(z)$ curves by $I(z)\propto \exp{(-z/\lambda)}$ as a function of sample bias $V$; $I_{stab}=V/(30$ G$\Omega$); symbols for measurements on graphene and Au(111) are indicated \protect{\cite{Zhang}} (courtesy of M. Crommie, UC Berkeley); (c) $dI/dV$ curves recorded on the same position of graphene deposited on SiO$_2$ at different $I_{\rm stab}$ as marked, $V_{\rm stab}=0.5$ V, $T=5$ K
\protect{\cite{Geringer2}}.}
\label{Fig4}
\end{figure}
First spectroscopic results have been published for graphene on SiC(0001), which exhibit a clear
Dirac cone dispersion by analyzing the wave length dependence of scattering states \cite{Stroscio}, but the
$dI/dV$ curves remain elusive \cite{Brar}.\\
On graphene flakes, Zhang et al. \cite{Zhang} observed a gap around the Fermi level
which appears to be independent from the back gate voltage moving the Dirac point through the Fermi level.
This gap of about 130 meV is shown in Fig. \ref{Fig4}(a). It has been related to the fact that tunneling into the graphene is preferentially to electronic states at the $\Gamma$-point, but not to the $K$-point, where the Dirac cone is located. Tunneling into the Dirac cone, thus, requires additional momentum, which, e.g., can be delivered by exciting a phonon. The lowest energy of a phonon with sufficient momentum is the out-of-plane accoustical phonon at the $K$ point, which, indeed, has an energy of about 65 meV \cite{Phonon} naturally explaining the gap of $\pm 65$ meV.
Additional evidence comes from the stronger decay of the tunneling current $I$ with tip-surface distance $z$ within the gap (Fig. \ref{Fig4}(b)), where direct tunneling into $K$-point electrons is required. A fit by the usual formula to the decay length $\lambda^{-1}  = 2\cdot \sqrt{2m\Phi/\hbar + k_{||}^2} $ ($m$: electron mass, $\phi=4.85$ eV: work function average tip/surface, $\hbar$: Planck's constant, $k_{||}$: wave vector parallel to the surface)  indeed nicely fits the data.\\
However, other authors \cite{Deshpande2a} did not observe the gap or do observe the gap
at low current only. The latter is shown in Fig. \ref{Fig4}(c), where a gap of 90 meV is observed  at stabilization currents below 100 pA, but closes at higher current \cite{Geringer2}. Also measurements of graphene on graphite did not observe the gap, but a linearly increasing $dI/dV(V)$ away from the Dirac point (see Fig. \ref{Fig7}(a)) \cite{Andrei2}. Thus, it is likely that details of the tip-sample geometry determine, if direct tunneling from the tip to $K$-point electrons is suppressed.\\
\subsection{Standing Waves and Potential Fluctuations}
\begin{figure*}[htb]
\centering{\includegraphics*[width=0.9\linewidth]{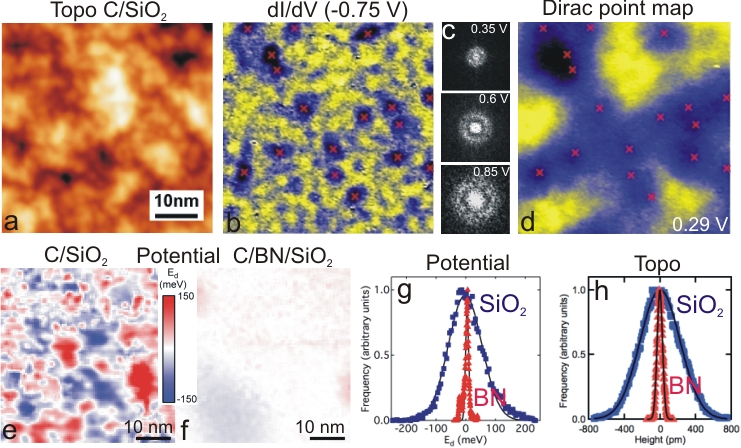}
\caption[]{(a) STM image of graphene on 285 nm SiO$_2$, $T=4.8$ K; (b) $dI/dV$ image of the same area, $V=-0.75$ V, $I=0.08$ nA, $V_{\rm gate}=60$ V; crosses mark the centers of standing waves; (c) Fourier transformation of $dI/dV$ images obtained at $V=0.35$ V, $I=0.05$ nA (top), $V=0.6$ V, $I=0.06$ nA (middle), and $V=0.85$ V, $I=0.07$ nA (bottom), $V_{\rm gate}=15$ V ;  (d) $dI/dV$ image of the same area as (a), (b), but recorded close to the Dirac point, $V=-0.29$ V, $I=0.025$ nA, $V_{\rm gate}=15$ V; crosses at positions as in (b) \protect{\cite{Zhang2}};
(e) Dirac point map deduced from the minimum in $dI/dV$ curves of graphene on 285 nm SiO$_2$, $T=4.5$ K; (f) same as (e) for graphene on 14 nm hexagonal BN on top of 285 nm SiO$_2$ ; (g) histogram of the Dirac point values $E_{\rm d}$ obtained in (e) and (f); (g) histogram of the height values obtained for graphene on BN/SiO$_2$ and SiO$_2$ \protect{\cite{Deshpande2}}; ((a)$-$(d): courtesy of M. Crommie, UC Berkeley, (e)$-$(h): courtesy of B. LeRoy, University of Arizona, Tuscon).}
\label{Fig5}}
\end{figure*}
STS can be used to map standing waves of Dirac electrons scattered by potential disturbances. Zhang et al. \cite{Zhang2} have found such standing waves on graphene flakes on SiO$_2$ in $dI/dV$ images as shown in Fig. \ref{Fig5}(b). The Fourier transformation of the real space data (Fig. \ref{Fig5}(c)) revealed circular structures indicating a dominating wave vector value $|\underline{k}|$ resulting from intracone scattering. Plotting this value as a function of bias voltage revealed a gapped linear $E(|\underline{k}|)$ dispersion in accordance with the gap described in Fig. \ref{Fig4}. But the steepness of the linear dispersion results in a Dirac velocity of $v_D\simeq 1.4\cdot10^6$ m/s, which is about 35 \% larger than expected \cite{Castro}. Since the authors
could change the gate voltage, they could exclude an influence of tip induced band bending \cite{Feenstra,Dombrowski} on this large value. The circular Fourier transformation pattern indicates intracone backscattering, which requires short range scatterers on monolayer graphene \cite{Ando}. A mapping of the lateral Dirac point distribution, performed as a dI/dV image slightly below the Dirac point (Fig. \ref{Fig5}(d)), revealed that the centers of the standing wave patterns induce a lowering of the Dirac point, which is interpreted as evidence of a charged donor character of the scatterers.\\
Standing wave patterns on bilayer material on SiC(0001) also exhibit contributions from intracone as well as from intercone scattering processes \cite{Stroscio} with the Fourier transformations of the intracone part revealing a linear dispersion gving a Dirac velocity of $1.0 \cdot 10^6$ m/s. Fourier transformations of standing wave patterns from monolayer graphene on SiC(0001) were interpreted as evidence for the absence of intracone backscattering \cite{Brihuega}.\\
The Dirac cone mapping in Fig. \ref{Fig5}(d) revealed potential fluctuations by about $\pm$ 30 meV on a 10$-$20 nm length scale. Xue et al. \cite{Deshpande2} observed a much larger fluctuation of $\pm 150$ meV on length scales of 3$-$10 nm as shown in Fig. \ref{Fig5}(e) again pointing to the
difference of nominally identically prepared samples. A strong reduction of potential fluctuations is achieved by putting an intermediate BN layer between the graphene and the SiO$_2$. This reduces the potential fluctuations to $\pm 5$ meV at a length scale of 50$-$100 nm as shown in Fig. \ref{Fig5}(f) \cite{Deshpande2}. The difference between the substrates is emphasized by comparing the histograms of potential values displayed in Fig. \ref{Fig5}(g). Interestingly, the difference of histograms is very similar to
the difference of histograms of the topographical corrugation of graphene on the two substrates. This is evident by comparing Fig. \ref{Fig5}(g) and (h). Notice that other authors found larger potential corrugations of $\pm 15$ meV on graphene on BN/SiO$_2$
\cite{Dekker}.

\subsection{Edge States}
\begin{figure}[htb]
\includegraphics[width=\linewidth]{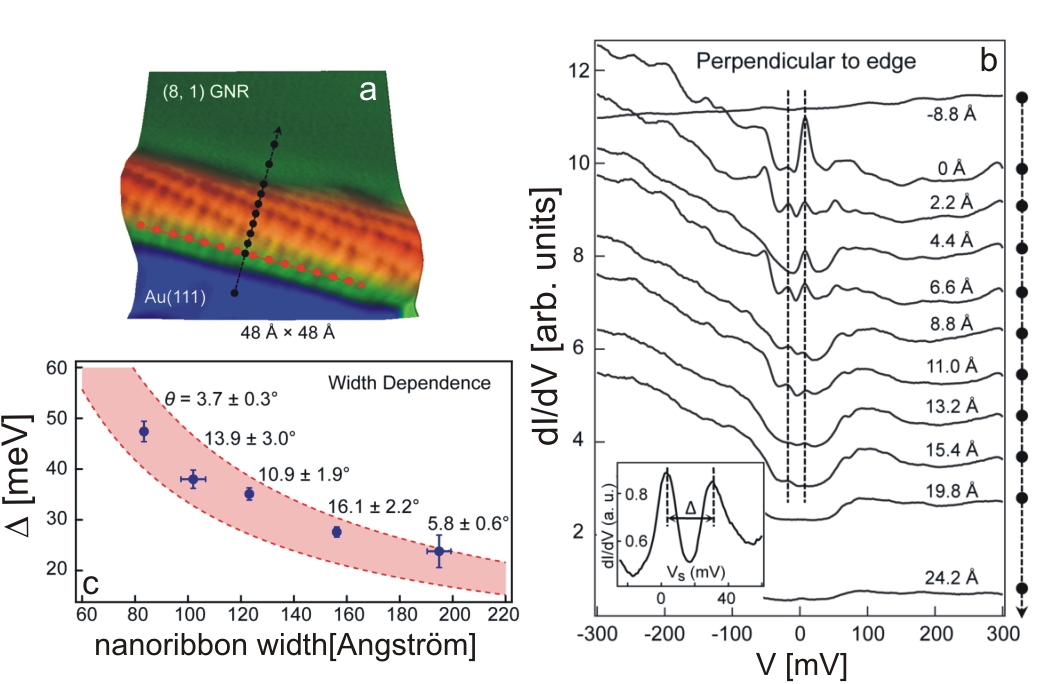}
\caption{(a) 3D representation of an STM image of a  graphene nanoribbon of width 19.5 nm having an (8,1) edge,  $V=0.3$ V, $I=0.06$ nA, $T=7$ K, the arrow with dots marks the positions where the spectra in (b) are recorded; (b) $dI/dV$ curves obtained on the nanoribbon shown in (a) at the dots marked; distance from the edge is indicated in $\rm \AA$ and peaks are highlighted by dashed lines; curves are offset for clarity; inset shows the spectrum at the edge of a smaller ribbon with the energy distance of peaks $\Delta$ marked, $V_{\rm stab}=0.15$ V, $I_{\rm stab}=0.05$ nA; (c) peak distance $\Delta$ as a function of nanoribbon width (symbols) in comparison with predictions from a Hubbard  model calculation (pink area); the angles are measured with respect to the zig-zag direction   \protect{\cite{Crommie}} (courtesy of M. Crommie, UC Berkeley).}
\label{Fig6}
\end{figure}
A very intriguing property of graphene is that unreconstructed zig-zag edges, which are singly terminated, should exhibit a magnetic edge state, which, moreover, is antiferromagnetically coupled to the zig-zag edges terminated by the other sublattice \cite{edgestate}. Interesting proposals, e.g. for tunable spin filters are based on that fact \cite{Son}. However, the existence of the edge states has been challenged by DFT calculations \cite{Mauri} revealing a thermodynamic instability of the edge states in realistic environments.
STS measurements have been performed on nanoribbons, which are produced by zipping calcinated nanotubes in an organic solution and subsequently using an ultrasonic bath \cite{Dai}. The nanoribbons are dispersed on a Au(111) surface. Figure \ref{Fig6}(a) shows an STM
image of such a nanoribbon. The edge area forms a little bump of about 4-5 $\rm \AA$ in height indicating bond formation of the edge with the substrate. From the atomic resolution on the bumped region, one can identify the edge geometry as (8,1) meaning that the edge consists of repeated segments of 8 zig-zag unit cells and a perpendicular armchair unit cell. The spectroscopy at the edge shown in Fig. \ref{Fig6}(b) reveals a double peak around the Fermi level in line with expectations from two magnetic edge states at the two ribbon edges, which are coupled antiferromagnetically to each other. The double peak disappears in the direction perpendicular to the edge after 2-3 unit cells. Figure \ref{Fig6}(c) shows the energy distance $\Delta$ of the peaks measured for nanoribbons of different widths. The value of $\Delta$ decays with width as expected from the antiferromagnetic coupling of the edges. The values for $\Delta$ are in quantitative correspondence with a Hubbard model using a hopping term $t=2.7$ eV and an on-site energy $U=1.35$ eV \cite{Crommie}.\\
This is good evidence that the edge state can survive even in solution, although a direct proof of its magnetic character is desirable.
\subsection{Landau levels}
\begin{figure*}[htb]
\centering{\includegraphics*[width=0.9\linewidth]{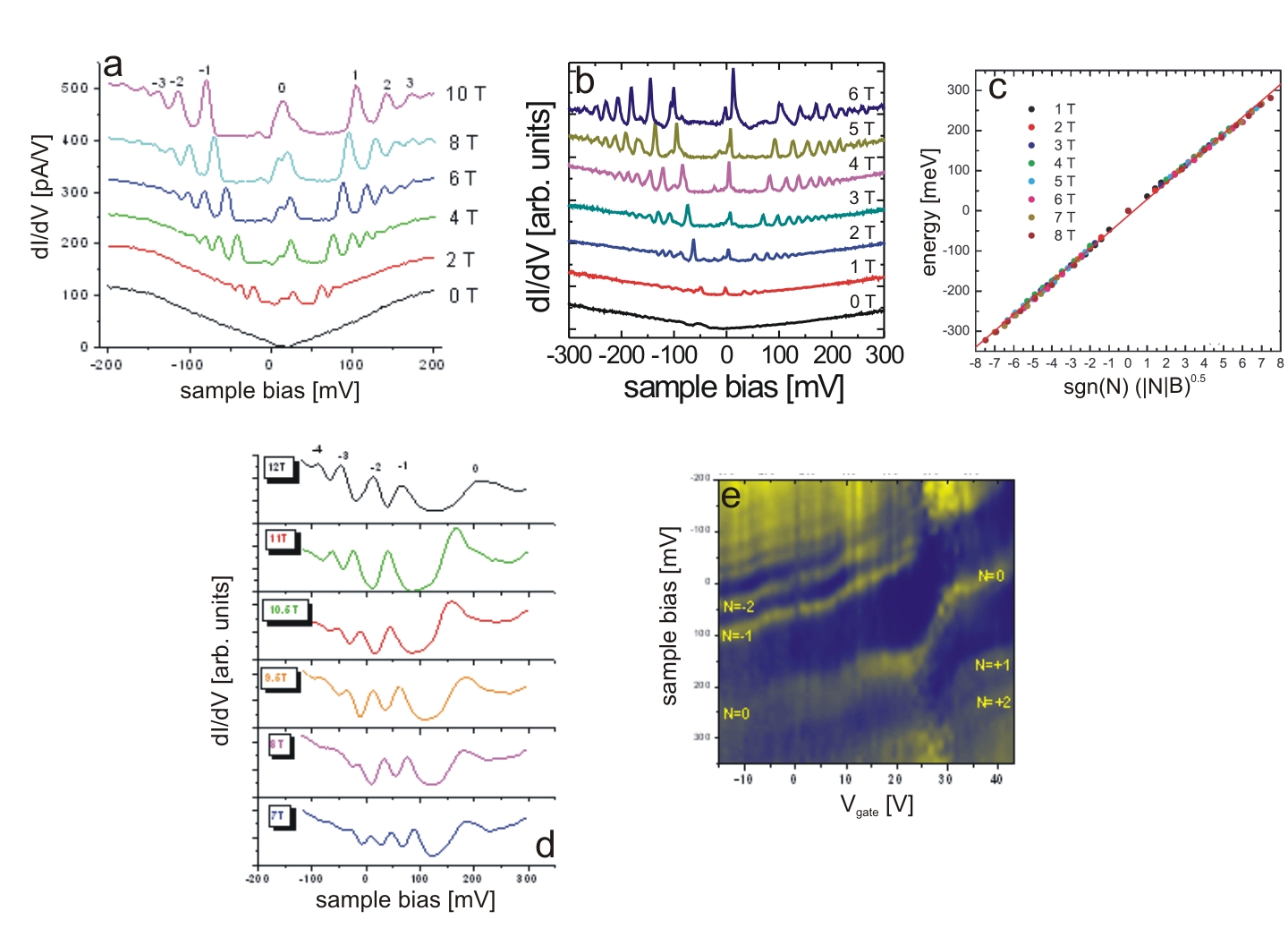}
\caption[]{(a) $dI/dV$ curves recorded on graphene on HOPG at different $B$ fields as marked, Landau level indices $N$ are marked at the peaks of the (10 T)-curve, $V_{\rm stab}=0.3$ V, $I_{\rm stab}=0.02$ nA, $T=4.4$ K \protect{\cite{Andrei2}}; (b) $dI/dV$ curves recorded on graphene on top of several graphene layers on SiC(000$\overline{1}$) at different $B$ fields as marked , $V_{\rm stab}=0.35$ V, $I_{\rm stab}=0.4$ nA, $T=4.3$ K \protect{\cite{Stroscio2}};  (c) observed peak energies from (b) as  a function of ${\rm sgn}(N)\sqrt{|N|B}$ ($N\in Z$) ; (d) $dI/dV$ curves recorded on graphene on chlorinated SiO$_2$ at different $B$ fields as marked, $V_{\rm stab}=0.3$ V, $I_{\rm stab}=0.02$ nA, $T=4.4$ K; (e) color plot of $dI/dV$ intensity as a function of sample bias $V$ and gate voltage $V_{\rm gate}$ at $B=12$ T; other parameters as in (d), Landau level index $N$ is marked \protect{\cite{Andrei3}} ((a), (d), (e): courtesy of E. Andrei, Rutgers University, (b), (c): courtesy of J. Stroscio, NIST Gaithersburg).}
\label{Fig7}}
\end{figure*}
A nice fingerprint for STS sensitivity are Landau levels, which appear in magnetic field $B$ due to the orbital quantization of electrons. In 2D systems, the quantization is complete leading to intricate transport effects as the integer quantum Hall effect \cite{Klitzing} or the fractional quantum Hall effect \cite{Stoermer}. Both effects have been observed on graphene \cite{Geim,Andrei}. The fractional quantum Hall effect, however, requires a low disorder potential, which has only been achieved, e.g., by freely suspending the graphene \cite{Andrei} or depositing it onto an intermediate layer of BN(0001) between a SiO$_2$ substrate and graphene \cite{Deshpande3}. If the disorder is reduced sufficiently, it is found that the thermal stability of the fractional plateaus is much
larger than for the usual GaAs 2D systems \cite{Andrei,Deshpande3}. This is expected, since the electrons are vertically quenched into a single layer, while GaAs 2D systems exhibit thicknesses of several nm. Thus, the electron-electron interaction at the same
electron density is significantly larger in graphene.\\
Landau levels detected by STS, which have previously been observed on semiconductor 2D systems \cite{Morgenstern}, have been first found for graphene layers on highly oriented pyrolytic graphite (HOPG) as shown in Fig. \ref{Fig7}(a) \cite{Andrei2}. The graphene layer has a slightly larger distance to the underlying graphene layer than within HOPG, which
appears to be sufficient to electronically decouple the graphene layer from the substrate. The fingerprint of the graphene Landau levels is its energy dependence $E_N = \pm v_{\rm D} \cdot \sqrt{2e\hbar\cdot |N|B}$  ($v_{\rm D}\simeq 10^6$ m/s: Dirac velocity, $e$: electron charge, $N\in Z$: Landau level index) \cite{Castro}. The peaks shown in Fig. \ref{Fig5}(a) indeed exhibit such a behavior \cite{Andrei2}. Graphene Landau levels have also been observed on SiC(000$\overline{1}$) \cite{Stroscio2}, which is the carbon terminated face. The sublimation method leads to thicker graphite layers, however, with individual graphene layers rotated with respect to each other \cite{rotateSiC}. It turned out that the rotation leads to an effective decoupling of the individual layers, resulting, e.g., in graphene cyclotron absorbtion already at $B=0.2$ T \cite{Berger2}. The corresponding Landau levels are shown in Fig. \ref{Fig7}(b) with the constituent $\sqrt{|N|B}$ dependence displayed in Fig. \ref{Fig7}(c)
\cite{Stroscio2}. The Landau levels have been used to map the disorder potential, which fluctuates on a length scale of about 50 nm by $\pm 5$ meV  \cite{Stroscio2} being very similar to the potential fluctuations found for graphene on BN (see Fig. \ref{Fig5}(f)) or on InAs 2D systems \cite{Morgenstern2}. First attempts to measure Landau levels on graphene flakes on SiO$_2$ were not successful \cite{Geringer2} because of the strong disorder potential which could fluctuate by $\pm 150$ meV (see Fig. \ref{Fig5}(e)). However, by chlorinating the substrate, the disorder is sufficiently reduced such that Landau levels are observed (see Fig. \ref{Fig7}d) \cite{Andrei3}. By applying the gate voltage as shown in Fig. \ref{Fig7}(e), one can nicely observe the jumps of Landau levels to the Fermi level, e.g. between $N=-1$ and $N=0$ around $V_{\rm gate}=30$ V.

\subsection{Drift states}
\begin{figure*}[htb]
\centering{\includegraphics*[width=0.7\linewidth]{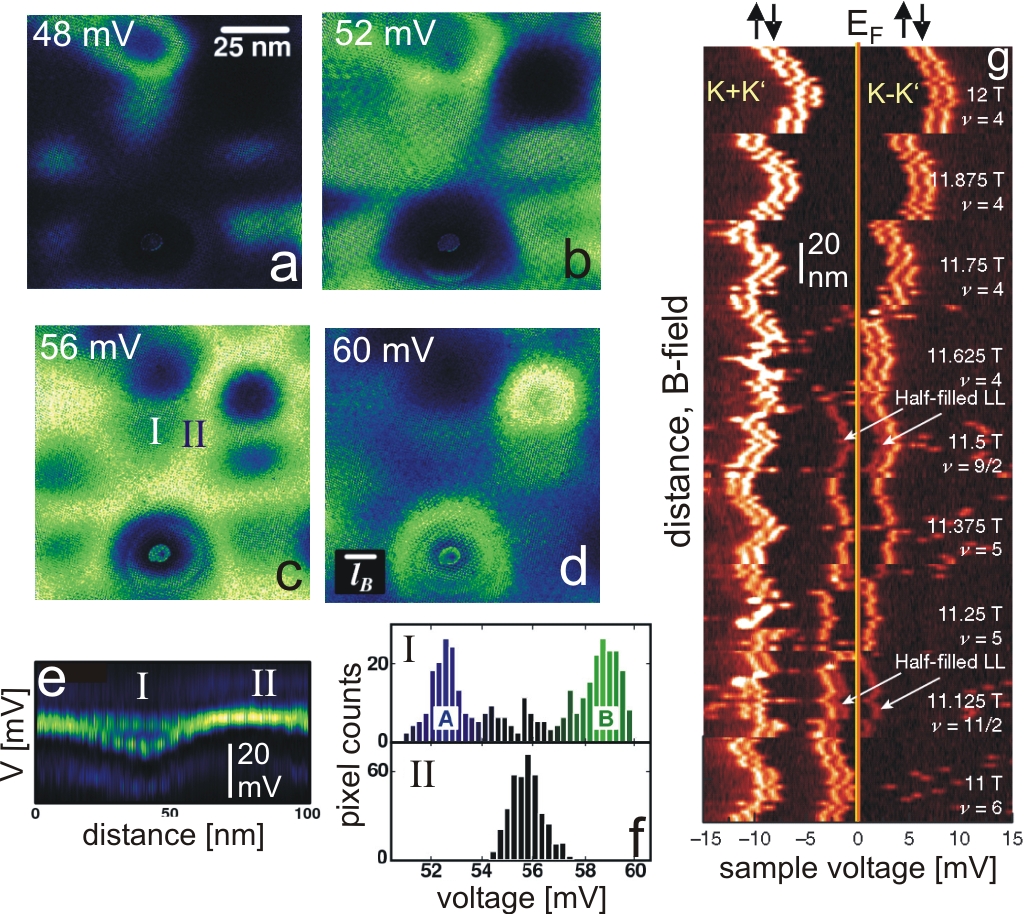}
\caption[]{(a)$-$(d) $dI/dV$ images recorded on graphene on SiC(000$\overline{1}$) at $B=8$ T at the voltages marked, $I=0.4$ nA, $T=4.3$ K, magnetic length $l_B=9$ nm is marked in (d), I and II in (c) mark areas similar to the ones, where the histograms in (f) have been obtained; (e) color plot of $dI/dV$ curves along a line showing the development of the $N=0$ Landau level, $V_{\rm stab}=0.35$ V, $I_{\rm stab}=0.4$ nA; I, II as in (c);
(f) histogram of peak values of dI/dV curves as displayed in (e) \protect{\cite{Miller}}; (g) color plots of $dI/dV$ curves around the Fermi level along a line displayed at different $B$ fields as marked, spin levels and (K,K') levels are indicated at the top part, $\nu$ is the local filling factor obtained from counting the number of peaks below the
Fermi level (hole Landau levels are not observed for an unknown reason), half filling factors are interpolated and marked, since contributing a third level, $V_{\rm stab}=0.25$ V, $I_{\rm stab}=0.2$ nA, $V_{\rm mod}=50$ $\rm \mu$V, $T=0.013$ K {\protect \cite{Song}} (courtesy of J. Stroscio, NIST Gaithersburg).}
\label{Fig8}}
\end{figure*}
The states corresponding to the Landau levels are called drift states. They meander along equipotential lines of
the disorder potential and exhibit a width of the magnetic length $l_B=\sqrt{\hbar/(eB)}$ \cite{Prange}. Most drift states are localized in potential valleys or at potential hills except of one state located
in the energetic center of the Landau level. This state traverses the whole potential landscape and, thus, is called the extended state. It is responsible for the quantum Hall transitions.
The transition from localized to extended states and back to localized states
has been imaged previously on InAs 2D systems \cite{Hashimoto}. It has also been found for the graphene layer on SiC(000$\overline{1}$) \cite{Miller}.
Fig. \ref{Fig8}(a)-(d) show dI/dV images at $B=8$ T taken at energies around the Landau level with index $N=0$. The magnetic length is marked in Fig. \ref{Fig8}(d) and indeed corresponds to the widths of the observed meandering structures.
A transition from localized states within the potential valleys ((a),(b)) via an extended state ((c)) to localized states around the potential hills ((d)) is apparent. Experiments on graphene flakes on SiO$_2$ did not show drift states. Instead, the strong disorder leads to strongly localized charge puddles, which exhibit Coulomb blockade peaks in STS experiments \cite{Jung}. Chlorinating the SiO$_2$ improves the disorder and indications of drift states become visible \cite{Andreip}.\\
A more detailed analysis of the drift states on SiC(000$\overline{1}$) reveals that a portion of the surface potential is given
by the lattice mismatch and rotation angle of the probed graphene layer with respect to
the underlying graphene layer. This implies a moir{\'e} pattern and a periodic potential with a lattice constant of about 70 nm \cite{Miller}. Interestingly, the $dI/dV$ curves exhibit a sublattice splitting in certain areas of this potential. Figure \ref{Fig8}(e) shows a line scan of $dI/dV$ curves. In the central area, the peak energy jumps between two values. Histograms of the peak values obtained in such areas are shown in the upper part of Fig. \ref{Fig8}(f). They reveal two dominating peak values being separated by about 7 meV. Other areas within the moir{\'e} structure do not exhibit a peak splitting as demonstrated by the histogram in the lower part of Fig. \ref{Fig8}(g). A plot of the peak splitting value as
a function of position reproduces the moir{\'e} pattern. The splitting is, thus, caused by the different stacking of the two graphene sublattices onto the underlying layer within these areas. Equivalently, one can say that the stacking of graphene to the sublayer locally lifts the valley degeneracy of the zero-order Landau level.\\
At very low temperature ($T=13$ mK), this system shows also spin splitting as demonstrated by the color plots in Fig. \ref{Fig8}(g) \cite{Song}. For example, at $B=12$ T, pairs of lines separated by the Zeeman energy of 1.4 meV are observed above and below the
Fermi level. Each pair of lines corresponds to one of the sublattices at the measurement position. The filing factor $\nu=4$ indicates that two more lines are observed below the Fermi level corresponding to the electron states of the $N=0$ Landau level. Thus, the four lines correspond to the $N=1$ Landau level. By decreasing the magnetic field, the degeneracy of each line decreases according to
$n_{\rm LL}=eB/h$ ($n_{\rm LL}$: number of states per m$^2$ in one spin- and valley-resolved Landau level). Thus, the levels above the Fermi level $E_{\rm F}$ must cross $E_{\rm F}$, if the electron density does not change. This crossing is visible in Fig. \ref{Fig8}(g). At the crossing point of the first spin level, the distance between the spin levels dramatically increases and it decreases again, if both spin levels have crossed $E_{\rm F}$. The effect is well known as exchange enhancement based on the fact that electrons with the same spin feel a weaker Coulomb repulsion than the ones with opposite spins \cite{exchange}. Moreover, a weak third line is observed at half-valued filling factor (marked by arrows in Fig. \ref{Fig8}(g)), which is interpreted as an interaction effect with the quantized 2D system of the underlying graphene layer. Notice that the exchange enhancement is the first tunable electron-electron interaction effect observed by STM on graphene.\\
Experiments of graphene flakes on BN/SiO$_2$ are under way and one should be curious if signatures of the fractional quantum Hall
plateaus \cite{Deshpande3} will be observed in STS.

\section{Conclusion}
Although  a number of very interesting STM and STS results on graphene have already been obtained ,
there are still multiple questions to be solved. One of the most interesting ones is the appearance of fractional quantum Hall states, where very little is known about. Other interesting aspects include the magnetism of the graphene edge state \cite{edgestate}, signatures of the so-called Zitterbewegung \cite{Zitterbewegung} or the proposed charge instability at highly charged defects \cite{Pereira}.
While most results so far took advantage of the fact that a 2D system is directly at the surface, future experiments might explore
the specifics of the Dirac nature of the particles. This might substantiate the idea to mimic elementary particle physics by small scale experiments and promises still an exciting time with STM on graphene. Most likely, also
the various applications now arising for graphene might pose questions to be solved on the nano-scale.

\begin{acknowledgement}
Helpful discussions with V. Geringer, M. Pratzer, M. Liebmann, F. Libisch, L. Wirtz, C. Stampfer, A. Georgi, T. Mashoff,
N. Freitag, M. I. Katsnelson, M. F. Crommie, W. G. Cullen, M. S. Fuhrer, J. Stroscio, and E. Y. Andrei as well as the figure supply of the
persons mentioned in the figure captions are gratefully acknowledged. Financial support by the German science foundation (DFG) via the projects Mo 858/11-1, Li 1050/2-1, and Mo 858/8-2 is appreciated.
\end{acknowledgement}

%

\begin{thebibliography}{[1]}

\bibitem{Geim}%
K. S. Novoselov, A. K. Geim, S. V. Morozov, D. Jiang, M. I. Katsnelson,
I. V. Grigorieva, S. V. Dubonos, and A. A. Firsov, Nature
{\bf 438}, 197 (2005);
K. S. Novoselov, A. K Geim, S. V. Morozov, D. Jiang, Y. Zhang, S. V. Dubonos, I. V. Grigorieva, and A. A. Firsov,
Science {\bf 306}, 666 (2004);
Y. B. Zhang , Y. W. Tan, H. L. Stoermer, and P. Kim, Nature {\bf 438}, 201 (2005).
\bibitem{Geim2}%
A. K. Geim and K. S. Novoselov, Nature Mat.  {\bf 6}, 183 (2007).
\bibitem{Castro}%
A. H. Castro Neto, F. Guinea, N. M. R. Peres, K. S. Novoselov, and A. K. Geim, Rev. Mod. Phys. {\bf 81}, 109 (2009).
\bibitem{Samsung}%
K. S. Kim, Y. Zhao, H. Jang, S. Y. Lee, J. M. Kim, K. S. Kim, J. H. Ahn, P. Kim, J. Y. Choi, and B. H. Hong,
Nature {\bf 457}, 706 (2009).
\bibitem{Avouris}%
Y. M. Lin, A. Valdes-Garcia, S. J. Han, D. B. Farmer, I. Meric, Y. N. Sun, Y. Q. Wu, C. Dimitrakopoulos, A. Grill, P. Avouris, and K. A. Jenkins, Science {\bf 322}, 1294 (2011);
Y. M. Lin, C. Dimitrakopoulos, K. A. Jenkins, D. B. Farmer, H. Y. Chiu, A. Grill, and P. Avouris, Science {\bf 327}, 662 (2010);
Y. Q. Wu, Y. M. Lin, A. A. Bol, K. A. Jenkins, F. N. Xia, D. B. Farmer, Y. Zhu, and P. Avouris, Nature {\bf 472}, (2011).
\bibitem{Ferrari}%
M. Liu, X. Yin, E. Ulin-Avila, B. Geng, T. Zentgraf, L. Ju, F. Wang, and X. Zhang, Nature {\bf 474}, 67 (2011);
F. Xia, T. Mueller, Y. Lin, A. Valdes-Garcia, and P. Avouris, Nature Nanotechnol. {\bf 4}, 838 (2009);
F. Bonaccorso, Z. Sun, T. Hasan, and A. C. Ferrari, Nature Photonics {\bf 4}, 611 (2010).
\bibitem{Kouwenhoven}%
S. Garaj, W. Hubbard, A. Reina, J. Kong, D. Branton, and J. A. Golovchenko, Nature {\bf 467}, 190 (2010);
S. K. Min, W. Y. Kim, Y. Cho, and K. S. Kim, Nature Nanotechnol. {\bf 6}, 162 (2011);
G. F. Schneider, S. W. Kowalczyk, V. E. Calado, G. Pandraud, H. W. Zandbergen, L. M. K. Vandersypen, and C. Dekker,
Nano Lett. {\bf 10}, 3163 (2010).
\bibitem{vanWees}%
N. Tombros, C. Jozsa, M. Popinciuc, H. T. Jonkman, and B. J. van Wees, Nature {\bf 448}, 571 (2007)
\bibitem{Beschoten}%
A. Avsar, T. Y. Yang, S. Bae, J. Balakrishnan, F. Volmer, M. Jaiswal, Z. Yi, S. R. Ali, G. Guentherodt, B. H. Hong, B. Beschoten, and B. Ozyilmaz, Nano Lett. {\bf 11}, 2363 (2011).
\bibitem{Ensslin}%
C. Stampfer, E. Schurtenberger, F. Molitor, J. Guttinger, T. Ihn, and K. Ensslin, Nano Lett. {\bf 8}, 2378 (2008);
F. Molitor, S. Droscher, J. Guttinger, A. Jacobsen, C. Stampfer, T. Ihn, and K. Ensslin, Appl. Phys. Lett. {\bf 94}, 222107 (2009).
\bibitem{Seyller}%
Th. Seyller, A. Bostwick, K. V. Emtsev, K. Horn, L. Ley, J. L. McChesney, T. Ohta, J. D. Riley, E. Rotenberg, and F. Speck,
phys. stat. sol. (b), {\bf 245}, 2437 (2008); T. Ohta, A, Bostwick, T. Seyller, K. Horn, and E. Rotenberg, Science {\bf 313}, 951 (2006); S. Y. Zhou, G. H. Gweon, A. V. Fedorov, P. N. First, W. A. de Herr, D. H. Lee, F. Guinea, A. H. Castro Neto, and A. Lanzara,
Nature Mat. {\bf 6}, 770 (2007).
\bibitem{Morgenstern3}
see e.g. J. Tersoff and D. R. Hamann, Phys. Rev. B {\bf 31}, 805 (1985); M. Morgenstern, D. Haude, V. Gudmundsson, C. Wittneven, R. Dombrowski, C. Steinebach, and R. Wiesendanger, J. Electr. Spectr. Rel. Phen. {\bf 109}, 127 (2000).
\bibitem{Chen}
C. J. Chen, J. Vac. Sci. Technol A {\bf 6}, 319 (1988); {\bf 9}, 44 (1991).
\bibitem{Mashoff}%
T. Mashoff, M. Pratzer, M. Liebmann, T. Echtermeyer, M. Lemme, and M. Morgenstern, Nano Lett. {\bf 10}, 461 (2010).
\bibitem{Meyer}%
K. Suenaga and M. Koshino, Nature {\bf468}, 1088 (2010);  J. C. Meyer, C. Kieselowski, R. Erni, M. D. Rossell, M. F. Crommie, and A. Zettl, Nano Lett. {\bf 8}, 3582 (2008); J. C. Meyer, C. O. Girit, M. F. Crommie, and A. Zettl, Nature {\bf 454}, 319  (2008).
\bibitem{Wallace}
P. R. Wallace, Phys. Rev. {\bf 71}, 622 (1947).
\bibitem{Peierls}
R. E. Peierls, Ann. I. H. Poincare {\bf 5}, 177 (1935); N. D. Mermin, Phys. Rev. {\bf 177}, 68 (1968).
\bibitem{Katsnelson}%
A. Fasolino, J. H. Los, and M. I. Katsnelson, Nature Mat. {\bf 6}, 858 (2007).
\bibitem{Meyer2}%
J. C. Meyer, A. K. Geim, M. I. Katsnelson, K. S. Novoselov, T. J. Booth, and S. Roth, Nature {\bf 446}, 60 (2007).
\bibitem{Locatelli}
A. Locatelli, K. R. Knox, D. Cvetko, T.O. Mentes, M.A. Nino, S. Wang, M. B. Yilmaz, P. Kim, R. M. Osgood Jr., and A. Morgante,
ACS Nano {\bf 4}, 4879 (2010).
\bibitem{Ando}%
T. Ando, T. Nakanishi, and R. Saito, J. Phys. Soc. Jpn. {\bf 67}, 2857 (1998).
\bibitem{Bolotin}
K. I. Bolotin, K. J. Sikes, J. Hone, H. L. St\"ormer, and P. Kim, Phys. Rev. Lett {\bf 101}, 096802 (2008).
\bibitem{Pereira}
V. M. Pereira, V. N. Kotov, and A. H. Castro-Neto. Phys. Rev. B {\bf 78}, 085101 (2008).
\bibitem{Zeitler}
K. S. Novoselov, Z. Jiang, Y. Zhang, S. V. Morozov, H. L. Stormer, U. Zeitler, J. C. Maan, G. S. Boebinger, P. Kim, and A. K. Geim,
Science {\bf 315}, 1379 (2007).
\bibitem{Prange}
R. Joynt and R. E. Prange, Phys. Rev. B {\bf 29}, 3303 (1984); R. E. Prange and S. M. Girvin, The Quantum Hall Effect (Springer, New York, 1987).
\bibitem{Andrei}%
X. Du, I. Skachko, F. Duerr, A. Luican, and E. Y. Andrei,
Nature {\bf 462}, 192 (2009); K. I. Bolotin, F. Ghahari, M. D. Shulman, H. L. Stoermer, and P. Kim, Nature {\bf 462}, 196 (2009).
\bibitem{Deshpande3}%
C. R. Dean, A.F. Young, P. Cadden-Zimansky, L. Wang, H. Ren, K. Watanabe, T. Taniguchi, P. Kim, J. Hone, and
K. L. Shepard, ArXiv1010.1179.
\bibitem{Girit}
C. Girit and A. Zettl, Appl. Phys. Lett. {\bf 91}, 193512 (2007).
\bibitem{Berger}%
C. Berger, Z. M. Song, X. B. Li, X. S. Wu, N. Brown, C. Naud, D. Mayou, T. B. Li, J. Hass, A. N. Marchenkov, E. H. Conrad, P. N. First, and W. A. de Heer, Science {\bf 312}, 1191 (2006)
\bibitem{Mallet}%
P. Mallet, F. Varchon, C. Naud, L. Magaud, C. Berger, and J. Y. Veuillen, Phys. Rev. B {\bf 76}, 041403 (2007).
\bibitem{Mallet3}
F. Varchon, R. Feng, J. Hass, X. Li, B. N. Nguyen, C. Naud, P. Mallet, J. Y. Veuillen, C. Berger, E. H. Conrad, and L. Magaud,
Phys. Rev. Lett. {\bf 99}, 126805 (2007); P. Lauffer, K. V. Emtsev, R. Graupner, Th. Seyller, L. Ley, S. A. Reshanov, and H. B. Weber, Phys. Rev. B {\bf 77}, 155426 (2008); G. M. Rutter, N. P. Guisinger, J. N. Crain, E. A. A. Jarvis, M. D. Stiles, T. Li, P. N. First, and J. A. Stroscio, Phys. Rev. B {\bf 76}, 235416 (2007).
\bibitem{Tikhonenko}%
F. V. Tikhonenko, D. W. Horsell, R. V. Gorbachev, and A. K. Savchenko, Phys. Rev. Lett. {\bf 100}, 056802 (2008).
\bibitem{Weber}%
D. Waldmann, J. Jobst, F. Speck, T. Seyller, M. Krieger, and H. B. Weber, Nature Mat. {\bf 10}, 357 (2011); A. Tzalenchuk, S. Lara-Avila, A. Kalalboukhov, S. Paolillo, M. Syvaejaervi, R. Yakimova, O. Kazakova, T. J. B. M. Janssen, V. Fal'ko, and S. Kubatkin, Nature Nanotechnol. {\bf 5}, 186 (2010).
\bibitem{Stolyarova}%
E. Stolyarova, K. T. Rim, S. M. Ryu, J. Maultzsch, P. Kim, L. E. Brus, T. F. Heinz, M. S. Hybertsen, and G. W. Flynn,  Proc, Natl. Acad. Sci. U.S.A. {\bf 104}, 9209 (2007).
\bibitem{Ishigami}%
M. Ishigami, J. H. Chen, W. G. Cullen, M. S. Fuhrer, and E. D. Williams, Nnao. Lett. {\bf 7}, 1643 (2007).
\bibitem{Geringer}%
V. Geringer, M. Liebmann, T. Echtermeyer, S. Runte, R. R\"uckkamp, M. Lemme, and M. Morgenstern, Phys. Rev. Lett. {\bf 102}, 076102 (2009).
\bibitem{Cullen}%
W. G. Cullen, M. Yamammoto, K. N. Burson, J. H. Chen, C. Lang, N. Li, M. S. Fuhrer, and E. D. Williams, Phys. Rev. Lett. {\bf 105}, 215504 (2010).
\bibitem{EModul}%
C. Lee, X. Wei, J. W.  Kysar, and J. Hone, Science {\bf 312}, 385 (2008).
\bibitem{Geim3}%
Y. W. Tan, Y. Zhang, K. Bolotin, Y. Zhao, S. Adam, E. H. Hwang, S. DasSarma, H. L. Stoermer, and P. Kim, Phys. Rev. Lett.
{\bf 99}, 246803 (2007).
\bibitem{Polini}%
M. Gibertini, A. Tomadin, M. Polini, A. Fasolino, and M. I. Katsnelson, Phys. Rev. B {\bf 81}, 125437 (2010).
\bibitem{Deshpande2}%
J. Xue, J. Sanchez-Yamagishi, D. Bulmash, P. Jacquod, A. Deshpande, K. Watanabe, T. Taniguchi, P. Jarillo-Herrero, and B. LeRoy,
Nature Mat. {\bf 10}, 282 (2011).
\bibitem{05Thz}%
N. Inui, K. Mochiji, and K. Moritani, Nanotechnology {\bf 19}, 505501 (2008).
\bibitem{Wiebe}%
S. H. Pan, E. W. Hudson, and J. C. Davies, Rev. Sci. Instr. {\bf 70}, 1459 (1999);
M. Kugler, Ch. Renner, O. Fischer, V. Mikheev, and G. Batey, Rev. Sci.
Instrum. {\bf 71}, 1475 (2000);
J. Wiebe, A. Wachowiak, F. Meier, D. Haude, T. Foster, M. Morgenstern, and  R. Wiesendanger, Rev. Sci. Instr. {\bf 75}, 4871 (2004).
\bibitem{Bachthold}%
A. Eichler, J. Moser, J. Chaste, M. Zdrojek, I. Wilson-Rae, and A. Bachthold, Nature Nanotechnol. {\bf 6}, 339 (2011).
\bibitem{Stroscio}%
G. M. Rutter, J. N. Crain, N. P. Guisinger, T. Li, P. N. First, and J. A. Stroscio, Science {\bf 317}, 219 (2007).
\bibitem{Brar}
V. W. Brar, Y. Zhang, Y. Yayon, T. Ohta, J. L. McChesney, A. Bostwick, E. Rotenberg, K. Horn, and M. F. Crommie, Appl. Phys. Lett.
{\bf 91}, 122102 (2007).
\bibitem{Zhang}%
Y. Zhang, V. W. Brar, F. Wang, C. Girit, Y. Yayon, M. Panlasigui, A. Zettl, and M. F. Crommie, Nature Phys. {\bf 4}, 627 (2008).
\bibitem{Phonon}%
M. Mohr, J. Maultzsch, E. Dobardzic, S. Reich, I. Milosevic, M. Damnjanovic, A. Bosak, M. Krisch, and C. Thomsen,
Phys. Rev. B {\bf 76}, 035439 (2007).
\bibitem{Deshpande2a} A. Deshpande, W. Bao, F. Miao, C. N. Lau, and B. J. LeRoy, Phys. Rev. B {\bf 79}, 205411 (2009).
\bibitem{Geringer2}%
V. Geringer, D. Subramaniam, A. K. Michel, B. Szafranek, D. Schall, A. Georgi, T. Mashoff, D. Neumaier, M. Liebmann, and M. Morgenstern, Appl. Phys. Lett. {\bf 96}, 082114 (2010)
\bibitem{Andrei2}%
G. Li, A. Luican, and E. Y. Andrei, Phys. Rev. Lett. {\bf 102}, 176804 (2009);
G. Li and E. Y. Andrei, Nature Phys. {\bf 3}, 623 (2007).
\bibitem{Zhang2}%
Y. Zhang, V. W. Brar, C. Girit, A. Zettl, and M. Crommie, Nature Phys. {\bf 5}, 722 (2009).
\bibitem{Feenstra}%
R. M. Feenstra and J. A. Stroscio, J. Vac. Sci. and Technol. B {\bf 5}, 923 (1987).
\bibitem{Dombrowski}%
R. Dombrowski, Chr. Steinebach, Ch. Wittneven, M. Morgenstern, and R. Wiesendanger, Phys. Rev. B {\bf 59}, 8043 (1999).
\bibitem{Brihuega}
I. Brihuega, P. Mallet, C. Bena, S. Bose, C. Michaelis, L. Vitali, F. Varchon, L. Magaud, K. Kern, and J. Y. Veuillen,
Phys. Rev. Lett., {\bf 101}, 206802 (2008).
\bibitem{Dekker}
R. Decker, Y. Wang, V. W. Brar, W. Regan, H. Z. Tsai, Q. Wu, W. Gannett, A. Zettl, and M. F. Crommie, Nano Lett. {\bf 11}, 2291
(2011).
\bibitem{edgestate}
K. Nakada, M. Fujita, G. Dresselhaus, and M. S. Dresselhaus, Phys. Rev. B {\bf 54}, 17954 (1996).
\bibitem{Son} Y. W. Son, M. L. Cohen, and S. G. Louie, Nature {\bf 444}, 347 (2006).
\bibitem{Mauri}
P. Koskinen, S. Malola, and H. H\"akkinen, Phys. Rev. Lett. {\bf 101}, 115502 (2008); T. Wassmann, A. P. Seitsonen, A. Marco Saitta, M. Lazzeri, and F. Mauri, Phys. Rev. Lett. {\bf 101} 096802 (2008).
\bibitem{Dai}
L. Y. Jiao, X. R. Wang, G. Dinakov, H. L. Wang and H. J. Dai, Nature Nanotechnol. {\bf 5}, 321 (2010).
\bibitem{Crommie}
C. Tao, L. Jiao, O. V. Yazyev, Y.-C. Chen, J. Feng, X. Zhang, R. B. Capaz, J. M. Tour, A. Zettl, S. G. Louie, H. Dai, and M. F. Crommie, Nature Phys. {\bf 7}, 616 (2011).
\bibitem{Klitzing}%
K. v. Klitzing, G. Dorda, and M. Pepper, Phys. Rev. Lett. {\bf 45}, 494 (1980).
\bibitem{Stoermer}%
D. C. Tsui, H. L. Stoermer, and A. C. Gossard, Phys. Rev. Lett. {\bf 48}, 1559 (1982);
\bibitem{Morgenstern}%
M. Morgenstern, J. Klijn, Chr. Meyer, and R. Wiesendanger, Phys. Rev. Lett. {\bf 90}, 056804 (2003).
\bibitem{Stroscio2}%
D. L. Miller, K. D. Kubista, G. M. Rutter, M. Ruan, W. A. de Heer, P. N. First, and J. A. Stroscio, Science {\bf 324}, 924 (2009).
\bibitem{rotateSiC}%
F. Varchon, P. Mallet, L. Magaud, and J. Y. Veuillen, Phys. Rev. B {\bf 77}, 165415 (2008).
\bibitem{Berger2}%
M. L. Sadowski, G. Martinez, M. Potemski, C. Berger, and W. de Heer,
Phys. Rev. Lett. {\bf 97}, 266405 (2006).
\bibitem{Morgenstern2}
M. Morgenstern, J. Klijn, Chr. Meyer, M. Getzlaff, R. Adelung, K. Rossnagel, L. Kipp, M. Skibowski, and R. Wiesendanger, Phys. Rev. Lett. {\bf 89}, 136806 (2002).
\bibitem{Andrei3}%
A. Luican, G. Li, and E. Y. Andrei, Phys. Rev. B {\bf 83}, 041405 (2011)
\bibitem{Hashimoto}
K. Hashimoto, C. Sohrmann, J. Wiebe, T. Inaoka, F. Meier, Y. Hirayama, R. A. R\"omer, R. Wiesendanger, and M. Morgenstern, Phys. Rev. Lett. {\bf 101}, 256802 (2008).
\bibitem{Miller}
D. L. Miller, K. D. Kubista, G. M. Rutter, M. Ruan, W. A. de Heer, M. Kindermann, P. N. First, and J. A. Stroscio,
Nature Phys. {\bf 6}, 811 (2010).
\bibitem{Jung}
S. Y. Jung, G. M. Rutter, N. N. Klimov, D. B. Newell, I. Calizo, A. R. Hight-Walker, N. B. Zhitenev, and J. A. Stroscio, Nature
Phys. {\bf 3}, 245 (2011).
\bibitem{Andreip}
E. Y. Andrei, private communication.
\bibitem{Song}
Y. J. Song, A. F. Otte, Y. Kuk, Y. Hu, D. B. Torance, P. N. First, W. A. de Heer, H. Min, S. Adam, M. D. Stiles, A. H. MacDonald, and J. A. Stroscio, Nature {\bf 467}, 185 (2010).
\bibitem{exchange}
T. Ando and Y. Uemura, J. Phys. Soc. Jpn. {\bf 37}, 1044 (1974).
\bibitem{Zitterbewegung}
M. I. Katsnelson, Eur. Phys. J. B {\bf 51}, 157 (2006).

\end{thebibliography}
%

\end{document}